\documentclass[a4paper,fleqn,usenatbib]{mnras}
\pdfoutput=1

\usepackage[pdftex]{graphicx,color}
\usepackage[T1]{fontenc}
\usepackage[utf8]{inputenc}
\usepackage{lmodern}

\definecolor{urlblue}{rgb}{0,0,0.9}
\definecolor{linkgreen}{rgb}{0,0.45,0}
\definecolor{linkorange}{rgb}{0.7,0.1,0.0}

\usepackage{amsmath, amssymb}
\usepackage{cuted}
\usepackage[english]{babel}
\usepackage{enumerate}
\usepackage[normalem]{ulem}
\usepackage{enumitem}
\usepackage{multirow}
\usepackage{booktabs}
\usepackage{textcase}
\usepackage{makecell}

\setlist[enumerate]{wide=0pt, widest=99,leftmargin=\parindent, labelsep=* } 

\definecolor{valecol}{rgb}{0,0.5, 1.}

\definecolor{davidcol}{rgb}{0.05,0.85, 0.05}

\definecolor{ziadcol}{rgb}{1.,.5, 0.0}

\graphicspath{{./figs/}}

\AtBeginDocument{\hypersetup{
linkcolor=linkgreen,
citecolor=linkorange,
urlcolor=urlblue}}

\title{The Copernican principle in light of the latest cosmological~data}

\author[Camarena, Marra, Sakr \& Clarkson]{
David Camarena,$^{1}$ Valerio Marra,$^{2,3,4}$ Ziad Sakr,$^{5}$ and Chris Clarkson$^{6,7,8}$
\\
$^{1}$PPGCosmo, Universidade Federal do Espírito Santo, 29075-910, Vitória, ES, Brazil\\
$^{2}$Núcleo de Astrofísica e Cosmologia \& Departamento de Física, Universidade Federal do Espírito Santo, 29075-910, Vitória, ES, Brazil\\
$^{3}$INAF -- Osservatorio Astronomico di Trieste, via Tiepolo 11, 34131, Trieste, Italy\\
$^{4}$IFPU -- Institute for Fundamental Physics of the Universe, via Beirut 2, 34151, Trieste, Italy\\
$^{5}$Université St Joseph; UR EGFEM, Faculty of Sciences, Beirut, Lebanon\\
$^{6}$School of Physics and Astronomy, Queen Mary University of London, UK\\
$^{7}$Department of Physics \& Astronomy, University of the Western Cape, Cape Town 7535, South Africa\\
$^{8}$Department of Mathematics \& Applied Mathematics, University of Cape Town, Cape Town 7701, South Africa
}

\begin{document}

\label{firstpage}
\pagerange{\pageref{firstpage}--\pageref{lastpage}}

\maketitle

\begin{abstract}
We pursue a program to confront observations with inhomogeneous extensions of the FLRW metric. The main idea is to test the Copernican principle rather than assuming it {\it a priori}.
We consider the $\Lambda$CDM model endowed with a spherical $\Lambda$LTB inhomogeneity around us, that is, we assume isotropy and test the hypothesis of homogeneity.
We confront the $\Lambda$LTB model with the latest available data from CMB, BAO, type Ia supernovae, local $H_0$, cosmic chronometers,  Compton y-distortion and kinetic Sunyaev–Zeldovich effect.
We find that these data can constrain tightly this extra inhomogeneity, almost to the cosmic variance level: on scales $\gtrsim 100$ Mpc structures can have a small non-Copernican effective contrast of just $\delta_L \sim 0.01$.
Furthermore, the constraints on the standard $\Lambda$CDM parameters are not weakened after marginalizing over the parameters that model the local structure, to which we assign ignorance priors.
In other words, dropping the Copernican principle assumption does not imply  worse constraints on the cosmological parameters. This positive result confirms that the present and future data can be meaningfully analyzed within the framework of inhomogeneous cosmology.
\end{abstract}


\section{Introduction}\label{sec:intro}

Cosmology studies the largest possible spatial and temporal scales of the observable  universe and, as a consequence, relies strongly on principles that can simplify our understanding of the spacetime.
Indeed, most observations are a collection of redshifted photons which are difficult to interpret  without a framework that can be used to disentangle temporal evolution from a possible spatial variation around us.
In order to make progress cosmologists have been assuming the Copernican principle, according to which we do not occupy a special location in the universe. In addition, if the universe is statistically isotropic, it then follows its statistical homogeneity, leading to the validity of the FLRW metric, the backbone of the standard cosmological model. By adopting the FLRW metric cosmologists made terrific progress in our understanding of the universe, providing a quantitative description of its evolution since the beginning of time and at all observable scales.

Cosmology now started mapping good fractions of the observable universe, soon producing  data at the rate of petabytes per year. This wealth of information may show that previously assumed hypotheses need to be relaxed, in particular the one of the FLRW metric.
The universe may indeed feature large-scale inhomogeneities and isotropies which cannot be explained by the standard model of cosmology.
While the Copernican principle may still be valid on much grander scales than the observable universe, it could well be discordant with our observations.
It follows that it is imperative to test the FLRW metric, the ultimate goal being to reconstruct the metric from observations \citep{Stebbins:2012vw}.

The FLRW metric can be tested through two complementary approaches: developing consistency tests and constraining inhomogeneous models.
The first approach aims at falsifying FLRW~\citep[see][for a review]{Clarkson:2012bg}, while the second at discovering features and structures beyond the standard model.

The second approach has been pursued by computing, for example, the fractal dimension in both two \citep{Alonso:2014xca, Goncalves:2017dzs} and three \citep{Scrimgeour:2012wt} dimensions using galaxy catalogs, showing a good agreement with the standard model.
Isotropy has been tested using the CMB \citep{Akrami:2014eta}, supernovae \citep[][]{Sun:2018cha,Zhao:2019azy,Krishnan:2021jmh}, compact radio sources \citep{Jackson:2012jt},  quasars \citep{Hirata:2009ar, Siewert:2020krp,Secrest:2020has}, galaxies \citep{Nadolny:2021hti} and clusters of galaxies \citep[][]{Migkas:2020fza,Migkas:2021zdo}.\footnote{Other ways to test inhomogeneity have been proposed, such as the time dependence  of  the  polarization  of  the CMB  photons  that  have  been inverse-Compton scattered by the hot gas in massive clusters of galaxies \citep{Jimenez:2019cll}.}

Here, we assume isotropy and test the hypothesis of homogeneity around us using the method proposed in \citet{Valkenburg:2012td}, that is, we test the validity of the Copernican principle.
We adopt the $\Lambda$LTB model \citep[see, e.g.][]{Marra:2010pg} which is basically the standard $\Lambda$CDM model endowed with a spherical over/underdensity.
The observer will sit at the center of the spherical structure. In other words, we neglect anisotropic degrees of freedom or, equivalently, average the observer's observations over angles.

We constrain the size and contrast of the spherical structure using the latest cosmological observations, and compare the result with the expectation from the Copernican prior---the probabilistic counterpart of the Copernican principle.
In order to consider the full likelihoods, we combine \texttt{MontePython} \citep{Audren:2012wb} for the MCMC exploration and likelihoods, \texttt{CLASS} \citep{Blas:2011rf} for the CMB computation  and \texttt{VoidDistances2020} \citep{Valkenburg:2011tm} for the $\Lambda$LTB metric functions via a wrapper that translates  the \texttt{MontePython} trial vector into an effective FLRW vector that is suitable for \texttt{CLASS}. We make publicly available the full \texttt{monteLLTB} pipeline at \href{https://github.com/davidcato/monteLLTB}{github.com/davidcato/monteLLTB}.
We consider the full Planck 2018 data \citep{Aghanim:2018eyx}, Pantheon supernovae \citep{Scolnic:2017caz}, the cosmic chronometer dataset \citep{Moresco:2015cya}, anisotropic and isotropic BAO distances \citep{Beutler:2011hx,Ross:2014qpa,Alam:2016hwk}, the Compton $y$-distortion \citep{Fixsen:1996nj}, the kinetic Sunyaev-Zeldovich effect \citep{Reichardt:2020jrr}, and the local constraint on $H_0$ via the local prior on the absolute magnitude $M_B$ of Type Ia supernovae \citep{Camarena:2019moy,Camarena:2021jlr}.

This paper is organized as follows. In Section~\ref{sec:LLTB} we briefly present the $\Lambda$LTB model, in Section~\ref{sec:obser} we discuss the observations that we consider  and how to confront them with $\Lambda$LTB, and in Section~\ref{cop} we introduce the Copernican prior.
We then show our results in Section~\ref{sec:results} and discuss them in Section~\ref{sec:discussion}. We conclude in Section~\ref{sec:conclusions}.

\section{A spherical inhomogeneous universe dominated by the cosmological constant}\label{sec:LLTB}

Following the notation of \citet{Biswas:2010xm}, the line element of the spherically symmetric LTB \citep{Lemaitre:1933gd,Tolman:1934za,Bondi:1947fta} metric can be written as 
\begin{equation}\label{eq:metric}
ds^2 = -dt^2 + \frac{R'^2(r,t)}{ 1+2r^2k(r)\tilde{M}^2} dr^2 + R^2(r,t) d\Omega \,,
\end{equation}
where $ d\Omega = d\theta^2 +\sin^2\theta d\phi^2$,  $\tilde{M}$ is an arbitrary mass scale and $k(r)$ is an arbitray function related to the curvature. The FLRW limit is reached through $R(r,t) \rightarrow a(t) r$ and $k(r) \rightarrow $ const, where $a(t)$ is the FLRW scale factor. Note that a prime denotes the partial derivative with respect to the radial coordinate $r$ and a dot will denote the partial derivative with respect to the time $t$.

Using Einstein's equations and an energy-momentum tensor containing the late-time $\Lambda$CDM components (matter and cosmological constant), we obtain the $\Lambda$LTB model, whose dynamics follows:
\begin{align}
\frac{\dot{R}^2(r,t)}{R^2(r,t)}  & = \frac{2m(r)}{R^3(r,t)} + \frac{2r^2k(r)\tilde{M}^2}{R^2(r,t)} +\frac{\Lambda}{3} \,, \label{eq:Hubble} \\
\rho_m(r,t) & =  \frac{m'(r)}{4 \pi G\, R'(r,t)R^2(r,t)} \,, \label{eq:rhom} 
\end{align}
where $\rho_m(r,t)$ is the energy density of matter, $G$ is the gravitational constant and $m(r)$ is the so-called Euclidean mass function \citep[][Appendix B]{Marra:2011zp}.

From the line element~\eqref{eq:metric}, we can note that the expansion of the universe is not only inhomogeneous but also anisotropic and, instead of a unique scalar factor, there are a transverse scale factor, $a_{\perp}(r,t) = R(r,t)/r $, and a longitudinal one, $a_{\parallel}(r,t) = R'(r,t)$. The corresponding expansion rates are then
\begin{align}
H_{\perp}(r,t) & = \frac{\dot{a}_{\perp}(r,t)}{a_{\perp}(r,t)} \,, \label{eq:HubbleT} \\
H_{\parallel}(r,t)  & = \frac{\dot{a}_{\parallel}(r,t)}{a_{\parallel}(r,t)} \,. \label{eq:HubbleL}
\end{align}
In addition, using the previous Friedmann-like equation, we can define the present-day density parameters of matter, curvature and cosmological constant as:
\begin{gather}
\Omega_m(r) = \frac{2 m(r)}{R^3(r,t_0) H_{\perp}^2(r,t_0)} \,, \label{eq:Omegam} \\ 
\Omega_k(r) = \frac{2 r^2 k(r) \tilde{M}^2}{R^2(r,t_0) H_{\perp}^2(r,t_0)} \,,  \label{eq:Omegak} \\
\Omega_\Lambda(r) = \frac{\Lambda}{3H_{\perp}^2(r,t_0)} \,.  \label{eq:OmegaL}
\end{gather}
For sake of simplicity hereafter we drop the subscript $\perp$ and simply use $a \equiv a_\perp$ and $H \equiv H_\perp$, unless stated otherwise. 
Combining equations~(\ref{eq:Omegam}--\ref{eq:OmegaL}) with equation~\eqref{eq:Hubble} it is possible to find the function $t(R,r)$, which is specified by the so-called Big Bang function $t_{BB}(r)$ \cite[see, e.g., eq.~(23) in][]{Valkenburg:2011tm}.

We have seen that the $\Lambda$LTB model is specified by three arbitrary functions: the mass function $m(r)$, the curvature profile $k(r)$, and the Big Bang function $t_{BB}(r)$.
One is but an expression of the gauge freedom which we fix here by setting $m(r) = 4 \pi \tilde{M}^2 r^3/3$~\citep{Biswas:2006ub,Biswas:2007gi}\footnote{This particular gauge excludes solutions with true vacuum over a finite $r$ interval \citep{Valkenburg:2011tm}.}.
The other two functions have physical meaning. By setting $t_{BB}(r)=\text{constant}$ one forbids decaying modes which would be in disagreement with the standard inflationary paradigm \citep{Silk:1977vv,Zibin:2008vj}.
Intuitively, this happens because the initial singularity would happen at different times for different shells so that large inhomogeneities would be present in the past.

One is then left with only one free function, $k(r)$, which then specifies the profile of the inhomogeneity, that is, its size and depth.
We adopt a compensated profile:
\begin{equation}\label{eq:kr}
k(r) = k_b + (k_c - k_b) P_3(r/r_B,0) \,, 
\end{equation}
where $k_b$ and $k_c$ are the curvature outside and at the center of the spherical inhomogeneity, respectively, $r_B$ is the comoving radius of the inhomogeneity and the function $P_n(x)$ is \citep{Valkenburg:2012td}:
\begin{align}
P_n (x) =  \left\lbrace 
\begin{array}{lll}
1-\exp \left [- \left (1- x \right)^n /x \right]  \phantom{ciao} & 0 \leq x < 1  \\ 
 0   & 1 \leq x .
\end{array}
\right. .
\end{align}

Profile~\eqref{eq:kr} ensures that the LTB and FLRW metrics match at the finite radius $r_B$ and also implies that there exists a radius $r_L < r_B$ at which the central over/underdense region makes the transition to the surrounding mass-compensating under/overdense region.
The use of a compensated profile guarantees that outside the LTB metric ($z>z_B$) one has exactly standard cosmology, particularly important for a consistent treatment of the CMB at $z \gg z_B$.
Also, a compensating over/underdense region is an expected feature of the standard large-scale structure: voids are surrounded by sheets and filaments, and superclusters by voids.

\begin{figure}
\centering 
\includegraphics[width=1.05\columnwidth]{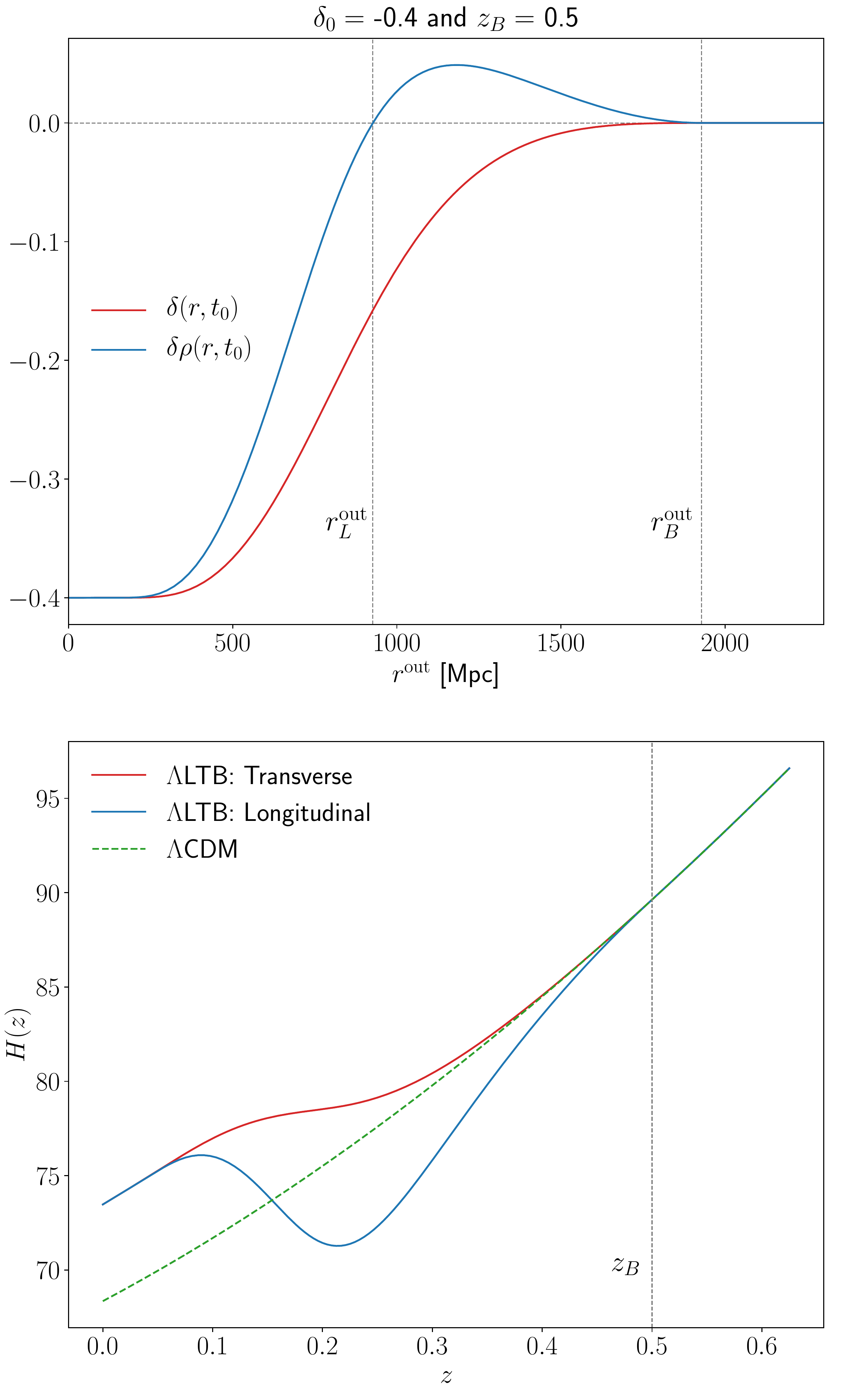}\caption{Top: The mass density contrast, $\delta(r,t_0)$, and matter density contrast, $\delta \rho (r,t_0) $, as functions of the FLRW radius of eq.~\eqref{rout}. The underdense region extends up to $r_L^{\rm out}$, at which $\delta \rho  = 0$. The compensating overdense region at $r_L^{\rm out} < r^{\rm out} < r_B^{\rm out}$ is necessary to have  $\delta (r_B,t_0) = 0$.
Bottom: The transverse and longitudinal Hubble rates of eqs.~\eqref{eq:HubbleT} and \eqref{eq:HubbleL} as a function of the redshift, as compared to the Hubble rate of the background $\Lambda$CDM model.
\label{fig:delta}}

\end{figure}

An example configuration is given in Figure~\ref{fig:delta}, where we show the matter density contrast:
\begin{align}
\delta \rho (r,t) =  \frac{\rho_m (r,t)}{ \rho_m (r_B,t)} - 1 \,,
\end{align}
and the (integrated) mass density contrast $\delta(r)$:
\begin{align}\label{eq:deltar}
\delta (r, t_0) &= \frac{4\pi \int_0^r d \bar r \, \delta \rho(\bar r,t_0)\,  R^2(\bar r,t_0) R'(\bar r,t_0) }{4\pi R^3( r,t_0) /3}  \\
&=  \frac{m(r)}{4\pi G R^3(r,t_0) /3 \, \rho_m^{\rm{out}}(t_0)} -1 = \frac{\Omega_m \, H_0^2}{\Omega_m^{\rm{out}}\, {H_{0}^{\rm{out}}}^2} -1 \,, \nonumber
\end{align}
where ``out'' denotes the corresponding FLRW quantities. 
Note that we are using a volume element without spatial curvature because it is the Euclidean mass that enters the Friedmann-like equation~\eqref{eq:Hubble}. The contribution of spatial curvature is, in any case, negligible for sub-horizon inhomogeneities \citep{Marra:2011ct}.
Note also that $\delta(r=0, t)=\delta \rho (r=0,t)$ and that $\delta(r=r_B,t)=\delta \rho (r=r_B, t)=0$.
In particular, the central contrast $\delta(r=0, t_0)=\delta_0$ is directly related to~$k_c$.

As said earlier, we fixed the freedom in the definition of $r$ via $m(r) = 4 \pi \tilde{M}^2 r^3/3$. This means that $r$ approximates the FLRW comoving coordinate only at initial time when the perturbation is small. At present time, the corresponding FLRW comoving coordinate is given by:
\begin{align} \label{rout}
r^{\rm out}= R(r,t_0)/a^{\rm out}(t_0) \,,
\end{align}
so that the FLRW and LTB physical distances coincide (neglecting again the negligible curvature contribution). Note that $r^{\rm out}_B=r_B$ because of the adopted matching condition.
Despite the fact that $r_B$ is the radius of the spherical inhomogeneity, the true scale of interest here is $r_L^{\rm out}$ since it defines the size of the central under/overdensity.

Our inhomogeneous universe is then specified by the inhomogeneous parameters and by the standard six $\Lambda$CDM parameters.
The latter are the normalized Hubble constant $h$, the baryon density $\Omega_b$, the cold dark matter density $\Omega_{cdm}$, the optical depth $\tau_{reio }$, the amplitude of the power spectrum $A_s$, and its tilt $n_s$.
Within our modeling, the inhomogeneity is specified by the parameters $z_B$ and $\delta_0$, where $z_B$ is the redshift corresponding to the radius $r_B$ of the spherical inhomogeneity, and $\delta_0$ is the matter density contrast at the center ($r=0$).

To improve the convergence of the Monte Carlo Markov Chain (MCMC), instead of $-1\le\delta_0<\infty$, we will use the following variable:
\begin{align} \label{deltatilde0}
\tilde \delta_0 =  \left\lbrace 
\begin{array}{lll}
\delta_0  \phantom{ciaociaociao} & \delta_0 \le 0  \\ 
\delta_0/(1+\delta_0)   & \delta_0 >0 
\end{array}
\right. ,
\end{align}
so that we can adopt a flat prior on $-1\le \tilde \delta_0\le1$. In the following we will omit the tilde.

\section{Observational probes}\label{sec:obser}

As said above, the observer sits at the center of the spherical structure, that is, we neglect anisotropic degrees of freedom or, equivalently, average the observer's observations over angles.
In order to confront with observations we then have to solve the corresponding geodesic equations:
\begin{align}
\frac{dt}{dz} & = -\frac{R'(r,t)}{(1+z)\dot{R}'(r,t)} \,, \label{eq:dtdz} \\
\frac{dr}{dz} & = -\frac{\sqrt{1+2r^2k(r)\tilde{M}^2}}{(1+z)\dot{R}'(r,t)}  \,. \label{eq:drdz} 
\end{align}
Although it is not possible to find analytically $R(r,t)$, one can use Carlson’s elliptic integrals \citep{1995NuAlg..10...13C} to accurately evaluate $t(R,t)$ from equation~\eqref{eq:Hubble}. Then, using numerical inversion, $R(r,t)$ can be precisely obtained \citep{Valkenburg:2011tm}. 
To perform this semi-analytic computation of the metric functions and LTB dynamic we use the \texttt{vd2020} code. We have embedded the \texttt{vd2020} code into the \texttt{montepython} code~\citep{Brinckmann:2018cvx,Audren:2012wb}, in order to take advantage of the likelihood structure and the MCMC sampler, resulting in the \texttt{monteLLTB} code, which is described in Appendix~\ref{ap:code}.

\subsection{Cosmic microwave background}\label{sub:CMB}

As we adopt a compensated profile which matches the FLRW metric at $z_B < 1$, the physics at (pre-)decoupling is as in the standard $\Lambda$CDM model.
Consequently, if we also assume the standard adiabatic power spectrum, changes on the CMB power spectrum are only produced by line-of-sight effects. More precisely, in comparison with a $\Lambda$CDM model, the spherical inhomogeneity  only changes the primary CMB spectrum via the late-time Integrated Sachs-Wolfe effect (ISW) and the angular distance to the last scattering surface.
It is important to stress that the choice of a standard power spectrum is \textit{a posteriori} justified, 
since observations will only allow radial inhomogeneities whose density contrast can be considered as a $\Lambda$CDM linear perturbation \citep{Valkenburg:2012td}. In this context, we moreover assume that the inhomogeneity does not change the late-time ISW effect as compared with $\Lambda$CDM.

Thus, an effective FLRW metric can be used to account for the changes produced in the CMB and it can be obtained through a rescaling of the background cosmology \citep{Zibin:2008vk,Marra:2010pg,Biswas:2010xm,Moss:2010jx}. 

Starting from the matching shell of coordinates  $\left\lbrace t_B \equiv t(r_B), r_B\right\rbrace$ and demanding the same angular distance in both the effective FLRW and the $\Lambda$LTB cosmology, we solve the geodesic equations of the $\Lambda$CDM background until $r=0$. This will give us the age of the effective FLRW cosmology, $t^{\rm{FLRW}}(r=0) = t_0^{\rm{eff}}$. In the same way, we can also obtain the boundary redshift as measured by an observer in the effective FLRW cosmology, $z_B^{\rm{eff}} = a^{\rm{FLRW}}(t_0^{\rm{eff}})/ a^{\rm{FLRW}}(t_B) -1$. We are then able to find the background  quantities of the effective FLRW model \citep[][eqs.~(3.6--13)]{Marra:2010pg}. Note that the non-background parameters, $A_s$, $n_s$ and $\tau_{\rm{reio}}$ will remain unchanged.

The CMB power spectrum of the effective FLRW model is computed through the \texttt{CLASS} code \citep{Blas:2011rf} (details in Appendix~\ref{ap:code}). We use the latest Planck observations for both high-$\ell$ and low-$\ell$ for the TT+TE+EE spectrum, available at \href{http://www.esa.int/Planck}{esa.int/Planck} \citep{Aghanim:2018eyx}.

Note that the impact of large-scale inhomogeneities on low-$\ell$ requires the challenging computation of  perturbations in an inhomogeneous background, mostly because of the complex contribution of the late ISW effect \citep{Tomita:2009wz,Clarkson:2010ej,Bolejko:2011jc}. However, as mentioned before, we assume that the late ISW effect is not modified by the spherical inhomogeneity because of the \textit{a posteriori}-small inhomogeneity contrast.
Nevertheless, in order to offer a robust analysis, we have also tested the impact of this assumption by performing an analysis without the low-$\ell$ Planck data, see Appendix~\ref{ap:lowl}.
\subsection{Type Ia Supernovae}\label{sub:SNe}

Supernovae Ia (SNe) are standardizable candles largely used in cosmology. Their apparent magnitudes, $m_B$, allow us to constrain cosmological models through the relation
\begin{equation}\label{eq:mb}
m_B(z) = 5\log_{10}\frac{d_L(z)}{1 \rm{Mpc}} + 25 + M_B \,,
\end{equation}
where $d_L$ is the luminosity distances and $M_B$ is the absolute magnitude. From the LTB metric \eqref{eq:metric}, one can note that the angular and luminosity distances, respectively, are:
\begin{align}
d_A(z) & = R(r(z),t(z)) \,, \label{eq:dA} \\
d_L(z)  & = (1+z)^2 R(r(z),t(z)) \,, \label{eq:dL}
\end{align}
where $t(z)$ and $r(z)$ are the solution to the geodesic equations \eqref{eq:dtdz} and \eqref{eq:drdz}. 

Here, we use the Pantheon dataset \citep{Scolnic:2017caz}, which contains a total of $1048$ supernovae in the redshift range  $0.01 < z < 2.3$. Unlike  previous SNe  datsets, as for instance JLA \citep{Betoule:2014frx}, the apparent magnitude $m_B$ of the Pantheon catalog already includes the correction due to stretch $x_1$, color $c$ and host-galaxy correction $\Delta_M$, leaving then $M_B$ as the only nuisance parameter. \footnote{An LTB analysis of the SDSS-II supernova dataset \citep{Kessler:2009} has shown that different light-curve fitters lead to different constraints on the LTB voids, especially on the size of inhomogeneity \citep{Bengochea:2014iha}. In this work, we do not explore this correlation between LTB parameters and light-curve fitter.}

We will consider both the full dataset and also the low-$z$ subset in the redshift range $0.023 \le z \le 0.15$ that is used to infer the Hubble constant via a cosmographic fit.

\subsection{Local prior}\label{sec:local}

In order to constrain very local scales it is important to include a prior on the Hubble constant.
As discussed in \citet{Camarena:2021jlr} \citep[see also][]{Benevento:2020fev,Efstathiou:2021ocp}, it is better to include the latter constraint via a prior on the absolute magnitude $M_B$ of Type Ia supernovae, removing the contribution from the cosmographic analysis that is adopted to fit for $H_0$.
The reasons are i) cosmography may fail when sudden low-redshift transitions are possible (especially relevant for the present case) and, in any case, its cosmographic parameters $q_0$ and $j_0$ will not agree with the ones adopted for the standard analysis (within LTB there is not a unique $H_0$ but instead $H_0(r) = H(r,t_0)$), ii) all supernovae are expected to share the same $M_B$, and iii) supernovae should not be double counted.
For more details, see \citet{Camarena:2021jlr}.

For the absolute magnitude of supernovae we adopt the effective gaussian prior $M_B =  -19.2334 \pm 0.0404$ from \cite{Camarena:2019moy}. This determination is obtained through a de-marginalization of the SH0ES determination in \citet{Reid:2019tiq}.

\subsection{Cosmic chronometers}\label{sub:cc}
Using spectroscopic techniques it is possible to determine the relative age between a pair of passively-evolving galaxies at different redshifts. Such differential age, along with the corresponding redshifts, can be used to determine the rate $dz/dt$ without  any assumptions  about cosmology. In an FLRW universe the rate $dz/dt$ simply corresponds to the Hubble parameter since $H(z) = \dot{a}/a$~\citep{Jimenez:2001gg}. In an LTB model, as it is clear from equations~\eqref{eq:dtdz} and~\eqref{eq:HubbleL}, the same rate corresponds to the radial Hubble parameter~$H_{\parallel}$. 

We use the dataset compiled in \citet[Table 4]{Moresco:2016mzx} to constrain $\Lambda$LTB. Such set contains 30 data points spanning the redshift range $0 < z < 2$ \citep{Moresco:2016mzx,Moresco:2012jh,Simon:2004tf,Stern:2009ep,Zhang:2012mp,Moresco:2015cya}.

\subsection{Baryonic Acoustic Oscillations}

At the drag epoch, $t_d$,  baryonic acoustic oscillations imprint the (comoving) sound horizon scale $r_d$ in the matter two point correlation function. Such scale can be used as a standard ruler, along both the longitudinal  and transverse directions.

Within  FLRW  both the longitudinal and transverse BAO scales are given by $\Delta z = l_d (1+z) H(z)$ and $ \Delta \theta = l_d/d_A(z)$, respectively, where $l_d = r_d/ (1+z)$ is the proper sound horizon. On the other hand, in a spherically inhomogeneous model, where the anisotropic expansion rates rule the dynamics, the BAO scales follow \citep{GarciaBellido:2008yq,Zibin:2008vk,Biswas:2010xm}:
\begin{align}
l_{\parallel}  & =  \frac{a_{\parallel}(r(z),t(z))}{a_{\parallel}(r(z),t_d)} \frac{r_d}{(1+z_d)} \,, \label{eq:lL} \\
l_{\perp} & = \frac{a(r(z),t(z))}{a(r(z),t_d)} \frac{r_d}{(1+z_d)} \,. \label{eq:lT}
\end{align}
leading then to
\begin{align}
\Delta z (z) & = l_{\parallel} (1+z) H_{\parallel}  \,, \label{eq:Deltaz} \\
\Delta \theta(z) & = \frac{l_{\perp}}{d_A(z)} \,, \label{eq:Deltatheta}
\end{align}
where $z_d$ is the redshift at the drag epoch, obtained using the effective FLRW model. 

Depending on the survey analysis, it is possible to detect both the radial, $\Delta z$, and angular, $\Delta \theta$, BAO scales or simply their isotropic combination
\begin{equation}\label{eq:dV}
d_V = r_d\left( \frac{z}{\Delta \theta^2 \Delta z} \right)^{1/3} \,.
\end{equation}
Here, we use both isotropic and anisotropic measurements coming from 6dFGS \citep{Beutler:2011hx}, SDSS-MGS \citep{Ross:2014qpa} and BOSS-DR12 \citep{Alam:2016hwk}. The isotropic measurements 6dFGS and SDSS-MGS allow us to assess low redshifts, $0.1$ and $0.15$, respectively, while the BOSS anisotropic data allow us to probe the redshifts $0.38$, $0.51$ and $0.61$.

BAO analyses make use of a fiducial cosmological model to analyze the observed redshifts and angles and so measure the transverse and longitudinal BAO peak positions. For a wide range of $w$CDM cosmologies, \citet{Carter:2019ulk} found no evidence for systematic errors in the measured BAO signal.
As the $\Lambda$LTB luminosity distance-redshift relation has a  phenomenology qualitatively similar to $w$CDM \citep[][]{Valkenburg:2013qwa} and the inhomogeneity contrast will be constrained to linear level by observations, the latter work suggests that the use of a fiducial $\Lambda$CDM model in the BAO analyses should not introduce a significant bias into our results.

\subsection{Compton y-distortion}

Reionized off-center structures can act as a mirror, scattering CMB photons within our past lightcone along our line-of-sight. This injects photons with different temperatures  and produces a spectral distortion, known as Compton y-distortion, of the CMB thermal black body spectrum. In the single-scattering and linear approximation, the spectral distortion produced by the off-center structure is \citep{Moss:2010jx,Caldwell:2007yu,Zibin:2011ma}: 
\begin{equation}
y = \frac{7}{10} \int^{z_{re}}_0 dz \frac{d\tau}{dz} \beta^2(z) \,, \label{eq:ydist} 
\end{equation}
where $z_{re}$ is the redshift of the reionization epoch, $\beta(z)$ is the dipole of the off-center structure and the time dependence of the optical depth $\tau$ is given by
\begin{equation}
\frac{d\tau}{dt} = \sigma_T f_b \left(1-\frac{Y_{He}}{2}\right) \frac{\rho_m (t)}{m_p} \,, \label{eq:dtaudt}
\end{equation}
where $\sigma_T$ is the Thompson cross section, $f_b$ is the baryon fraction, $Y_{He}$ is the helium mas fraction and $m_p$ is the proton mass.
Note that equation~\eqref{eq:ydist} assumes that the dominant contribution to the $y$-distortion is given by the dipole, neglecting the  higher multipoles. 

In $\Lambda$CDM, the dipole $\beta(z)$ is produced by peculiar velocities, that is, by perturbations. On the other hand, the very $\Lambda$LTB background produces a dipole for  off-center structures.
In fact, we can rougly approximate $\beta(z) \simeq  D \delta H$, where $D$ is some proper distance \citep{Alnes:2006pf}. Here, in order to provide an accurate estimation of $y$-Compton distortion, we compute $\beta(z)$ following the procedure stated in \cite{GarciaBellido:2008gd}. 
First, one identifies the  redshift of the off-center structure, $z$, then starting at coordinate $\left\lbrace t(z),r(z)\right\rbrace$, one solves the outgoing and ingoing geodesic equations to the surface of last scattering obtain $z_-$ and $z_+$, respectively. Then, considering that the temperature of CMB scales according to $T \propto 1/z$, the dipole in the light-cone is given by $\beta(z) = (z_+ - z_-)/(2+ z_+ +z_-)$ \cite[see Figure~1 in][]{GarciaBellido:2008gd}.

The $y$-Compton spectral distortion provides an interesting way to extract cosmological information that could be even useful to improve our understanding of the standard $\Lambda$CDM model \citep[see for instance][]{Lucca:2019rxf}. However, the current measurement  is not precise enough to provide any statistically significant constraint on $\Lambda$CDM. In fact, the only available measurement comes from the COBE-FIRAS satellite~\citep{Fixsen:1996nj}, which sets an upper limit at $2\sigma$ given by $y < 1.5 \times 10^{-5}$. Although this upper limit does not offer major information regarding  the $\Lambda$CDM paradigm, it nevertheless directly constrains the dipole, $\beta(z)$, and so spherical inhomogeneity.

\subsection{The kinetic Sunyaev–Zeldovich effect}\label{sec:kSZ}
The existence of a dipole for off-center structures also produces  anisotropies in the CMB spectrum via the kSZ effect. Generated by the inverse Compton scattering of low-energy photons with high-energy electrons, the kSZ effect is a powerful observable  to constrain inhomogeneous models \citep{GarciaBellido:2008gd,Zhang:2010fa,Moss:2011ze,Bull:2011wi}. 

Here, we will consider the so-called linear kSZ effect \citep{Zhang:2010fa}. Using the Limber approximation and considering the effect due to all free electrons in the reionized universe we compute the linear kSZ effect as \citep{Moss:2011ze}:
\begin{equation}\label{eq:kSZ}
C_\ell^{\text{kSZ}} \simeq  \frac{16\pi^2}{(2\ell+1)^3} \int_0^{r_\text{re}} dr \, r \left[\beta(r) \frac{d\tau}{dr}  \right]^2 \Delta_m^2 \,,
\end{equation}
where $r_{\text{re}}$ is the radial coordinate at $z_{\text{re}}$ and the nonlinear dimensionless matter power spectrum depends on $r$ according to:
\begin{align}
\Delta_m^2= \Delta_m^2 \Big (  \big( \bar k= \frac{2 \ell+1}{2 r} \big) \times \Xi, \, z(r) \Big) \,,
\end{align}
where $\Xi$ is introduced in order to correct for the LTB anisotropic expansion:
\begin{equation}\label{eq:hatk}
\Xi = \left(\frac{1+\overline{z}}{1+z}\right) \left[ \frac{a^2(\overline{t},r(z))}{a^2(t(z),r(z))} \frac{a_\parallel(\overline{t},r(z))}{a_\parallel(t(z),r(z))} \right]^{1/3} \,.
\end{equation}

Indeed, let us consider a comoving wavenumber $\bar k$ at an early-enough time $\bar t$ at which, thanks to the absence of decaying modes, the metric is close to FLRW and a harmonic decomposition of the temperature perturbations is possible. Because of the subsequent anisotropic expansion, the proper mode $\bar k/ \bar a$ is stretched differently along the longitudinal and transverse direction:
\begin{align}
\frac{k_{\left\lbrace \parallel,\perp \right\rbrace}}{a} & = \frac{\overline{k}}{\overline a} \, \frac{a_{\left\lbrace \parallel,\perp \right\rbrace}(\overline{t},r)}{a_{\left\lbrace \parallel,\perp \right\rbrace}(t,r)} \,.
\end{align}
As we need to feed a single wavenumber to the standard power spectrum, we will then consider, in analogy to the BAO scales, the isotropic wave number $\left[ k_{\perp}^2(z) k_{\parallel}(z) \right]^{1/3} $, justifying the previous equations.

We constrain  spherical inhomogeneity using the first kSZ measurement at more than $3\sigma$ given by $D^\text{obs}_{3000} = 3.0 \pm 1.0 \ \mu$K \citep{Reichardt:2020jrr}, where $ 2 \pi D_\ell = \ell (\ell +1 ) C_\ell$. We compute the non-linear power spectrum using the HALOFIT model \citep{Smith:2002dz} and considering the background FLRW cosmology. Note that, because of linear perturbations and peculiar velocities, the $\Lambda$CDM background also contributes to the kSZ effect, i.e., the kSZ effect does not disappear when $z_B, \delta_0 \rightarrow 0$. We take into account this $\Lambda$CDM contribution using the patchy and homogenous parameterizations \citep{Calabrese:2014gwa}:
\begin{align}\label{eq:backSZ}
\text{h-A}_\text{kSZ} & = 1.65\left(\frac{\sigma_8}{0.8}\right)^{4.46} \,, \\
\text{p-A}_\text{kSZ} & = 2.03\left[\frac{(1+z_{re})}{11}-0.22 \right] \left(\frac{\Delta z_{re}}{1.05}\right)^{0.51} \,,
\end{align}
where $\Delta z_{re} = z(x_i=25 \%) - z(x_i=75 \%)$ is the duration of reionization and $x_i$ is the ionization fraction of hydrogen. We compute $x_i$ using the $\tanh$ model \citep{Lewis:2008wr}.

It is worth stressing that our implementation of the kSZ effect is not free of ambiguities and is based on the {\it a posteriori} result that observations constrain the $\Lambda$LTB inhomogeneity to an almost linear perturbation of $\Lambda$CDM. A fully consistent treatment of kSZ requires the not-yet available understanding of the growth of matter perturbations in an inhomogeneous background.

\section{Copernican prior} \label{cop}

If the Copernican principle is valid, then the perturbations inferred from CMB observations should describe the early universe at any point and, in particular, also at our observing position.
It follows then that we can use CMB summary statistics such as the power spectrum to translate the Copernican principle into its statistical counterpart, the ``Copernican prior.'' 
Specifically, the Copernican prior enforces the requirement that \textit{local} inhomogeneities -- parametrized by $z_B$ and $\delta_0$ within our construction -- must agree with  the power spectrum as predicted by the CMB \citep{Valkenburg:2012td}.

To build the Copernican prior, we start by assuming that the density contrast, $\delta$, is a Gaussian field with a vanishing mean.
Under the assumption of a spherical inhomogeneity, we compute the variance of $\delta$ through the standard mean square estimator
\begin{equation} \label{eq:sigmar}
\sigma^2(r) = \int_0^{\infty} \frac{dk}{k} \Delta^2_{m0}(k) \left[3 \frac{j_1(r \, k)}{r \, k} \right]^2 \,,
\end{equation}
where $\Delta_{m0}(k)$ is the standard dimensionless power spectrum today and $j_1$ is the spherical Bessel function of the first kind. We adopted the linear power spectrum as we are  interested on large scales ($\gtrsim 20$Mpc) at which the nonlinearities have a negligible impact on our results.
The radius $r$ is the size of the inhomogeneity that the Copernican prior will constrain. As we are considering a compensated profile, we cannot compute the likelihood of having a given perturbation on the scale $r_B$ as it is $\delta(r_B,t) = 0$ by construction (see Fig.~\ref{fig:delta}). Rather we must use the scale $r_L$ of the actual under/overdensity.
Thus, the Copernican prior is defined as 
\begin{equation} \label{eq:CP}
\mathcal{P}(\delta_0,z_B) \propto \exp{\left[ -\frac{1}{2} \frac{\delta^2(r_L(\delta_0,z_B),t_0)}{\sigma^2(r_L^{\rm out}(\delta_0,z_B))}  \right]} \,,
\end{equation}
where $\delta(r,t_0)$ is given in equation~\eqref{eq:deltar}, the function $r_L(\delta_0,z_B)$ gives the radius of the central under/overdensity given the central contrast $\delta_0$ and the redshift of the inhomogeneous patch $z_B$, and the FLRW radius $r^{\rm out}$ is defined in equation~\eqref{rout}.
We will then compare the observational constraints on $\Lambda$LTB with the ones from the Copernican prior convolved with the CMB likelihood:
\begin{equation} \label{PCMB}
  P( \delta_0 , z_B)  =  \int dp_i \mathcal{P}(\delta_0,z_B) \, \mathcal{L}_{\rm CMB}(p_i,\delta_0,z_B) \,,
\end{equation}
where $p_i$ denote the standard $\Lambda$CDM parameters and $\mathcal{L}_{\rm CMB}$ is the CMB likelihood of Section~\ref{sub:CMB}.
Therefore, $ P$ is the probability distribution of $\delta_0$ and $z_B$, given the initial conditions obtained from the CMB and their uncertainty, which, under the Copernican principle, describe matter perturbations around us.

Finally, the above prior differs from the one adopted in \citet{Valkenburg:2012td} on three aspects.
First, we removed the normalization factor $(\sigma_L \sqrt{2 \pi})^{-1}$, since it has the effect of weighting differently different values of $z_B$, while instead the Copernican prior should penalize in the same way fluctuation at any $z_B$.
Second, \citet{Valkenburg:2012td} uses the relativistic mass to compute $\delta (r)$ while we adopt the Euclidean mass as in~\eqref{eq:deltar}.
Quantitatively the difference is small and the Euclidean mass definition can be better compared with equation~\eqref{eq:sigmar}.
Lastly, we correct for the LTB gauge by using $r^{\rm out}$.

\section{Results}\label{sec:results}

\begin{figure*}
\centering 
\includegraphics[width=\textwidth]{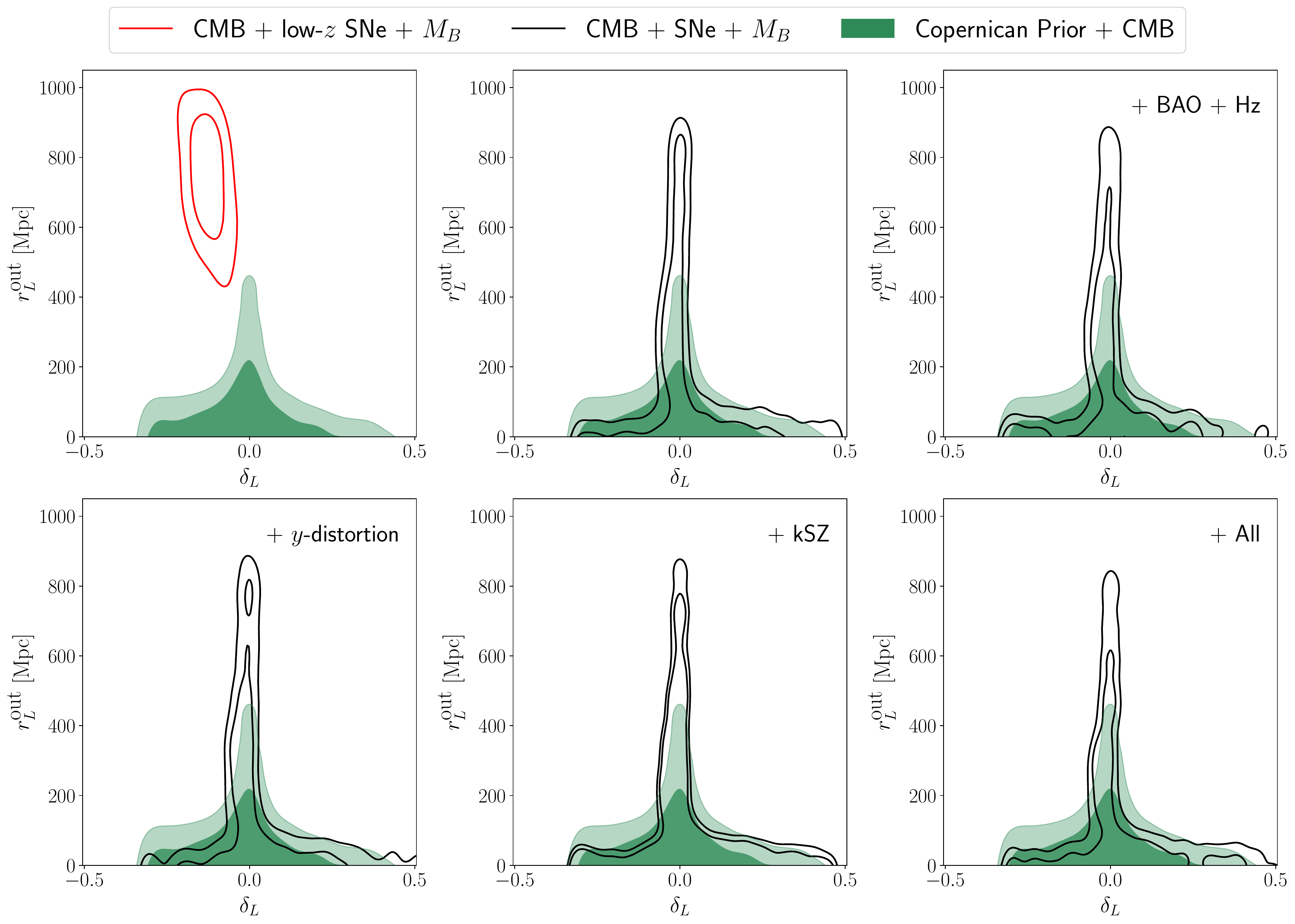} 
\caption{Marginalized constraints on the effective contrast $\delta_L$ and size $r_L^{\rm out}$ of the $\Lambda$LTB inhomogeneity at 68\% and 95\% confidence level. The empty contours show the constraints from the corresponding combination of observables.
The green area shows the region of the parameter space that is allowed by the standard model, here represented via the Copernican prior convolved with the CMB likelihood.}
\label{fig:LLTB-constraints}
\end{figure*}

\begin{figure}
\centering 
\includegraphics[trim={.7cm .7cm .7cm .7cm},clip, width=\columnwidth]{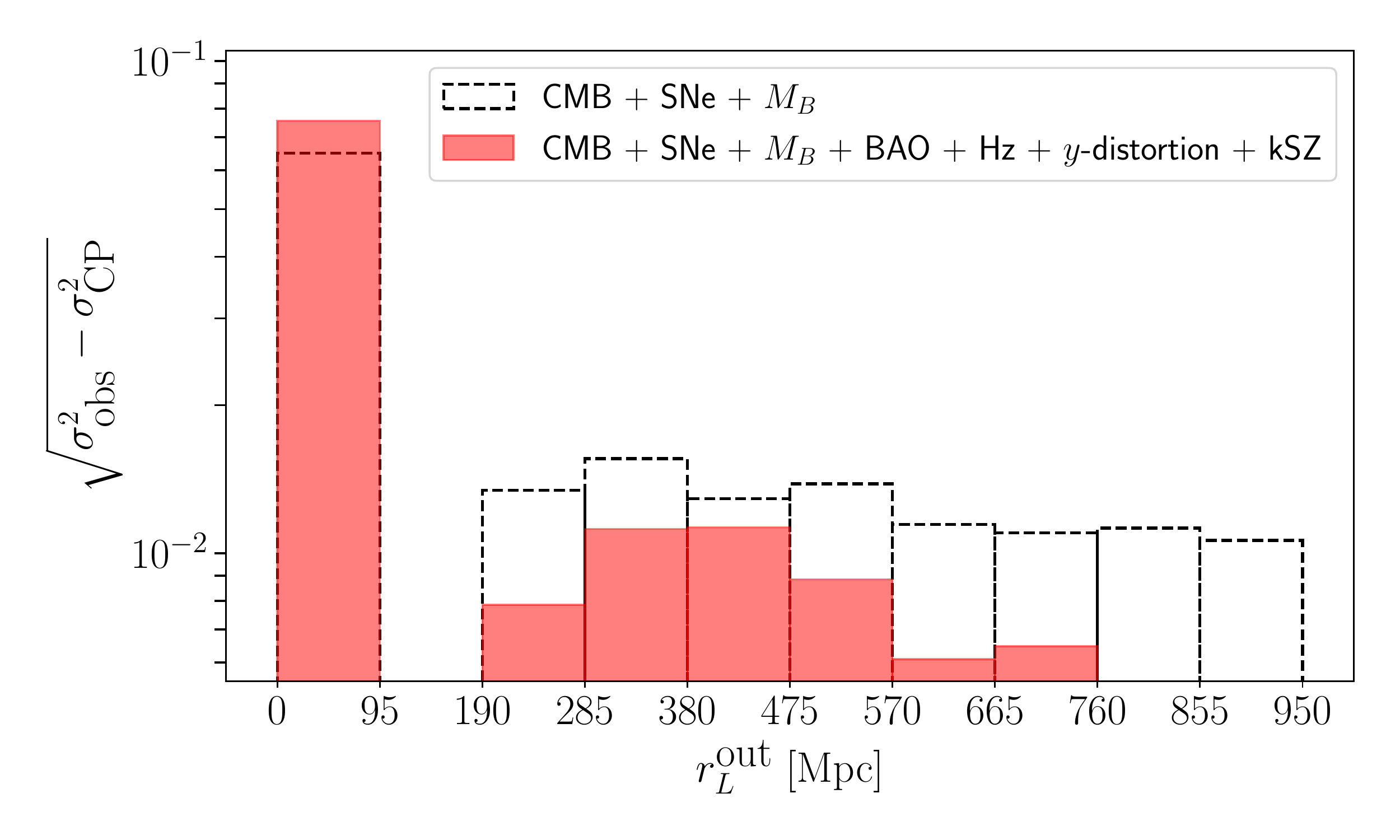} 
\caption{
Effective contrast beyond what is allowed by the Copernican principle (CP) as a function of the effective size $r_L^{\rm out}$ of the $\Lambda$LTB inhomogeneity.
One can see that non-Copernican  structures can have a small extra effective contrast of just $\delta_L \sim 0.01$.
}
\label{fig:excess}
\end{figure}

As mentioned earlier, we have performed the data analyses using the \texttt{monteLLTB} code (see Appendix~\ref{ap:code}). We use the Gelman-Rubin diagnostic \citep[$R$,][]{Gelman:1992zz} to evaluate the convergence of the Markov-chain Monte Carlo analysis. Explicitly, we demand chains with $(R-1) \lesssim 0.05$ for the inhomogeneity parameters $\delta_0$ and $z_B$. This leads to $\Lambda$CDM parameters with a convergence of $(R-1) \sim \mathcal{O}(10^{-3})$.
Most of the plots showed in this section have been produced using \texttt{getdist} \citep{Lewis:2019xzd}.

\subsection{Constraints on the inhomogeneity}

Figure~\ref{fig:LLTB-constraints} shows the marginalized constraints on the comoving size $r_L^{\rm out}$ and integrated mass contrast  $\delta_L=\delta(r_L, t_0)$ of the $\Lambda$LTB inhomogeneity for various combinations of observables.%
\footnote{As for $\delta_0$, we are actually showing $\tilde \delta_L$ instead of $\delta_L$,  as explained after eq.~\eqref{deltatilde0}.}
Also shown are the constraints from the Copernican prior convolved with the CMB likelihood of equation~\eqref{PCMB}, that is, the region of the parameter space that is allowed within the standard model of cosmology.
It is clear that only linear non-Copernican structures are allowed at larger radii once all the observables are considered, while for smaller sizes the Copernican principle (CP) is confirmed and, actually, observations start to map the local structure.

In order to better see this, we show in Figure~\ref{fig:excess} the effective contrast beyond what is allowed by the Copernican principle as a function of the effective size $r_L^{\rm out}$. We define this non-Copernican $\delta_L$ as $\sqrt{\sigma^2_{\rm obs}-\sigma^2_{\rm CP}}$ within the corresponding $r_L^{\rm out}$ bin, where $\sigma^2_{\rm obs}$ and $\sigma^2_{\rm CP}$ are the variances of $\delta_L$ relative to the empty and green contours of Figure~\ref{fig:LLTB-constraints}, respectively.
Figure~\ref{fig:excess} shows that structures can have a small extra effective contrast of just $\delta_L \sim 0.01$.%
\footnote{Note that, because of the non-Gaussian nature of the posterior, it is not straightforward to compare Figure~\ref{fig:excess} with Figure~\ref{fig:LLTB-constraints}.}

\subsection{Constraints on the the standard model parameters}

\begin{figure*}
\centering 
\includegraphics[width=0.9\textwidth]{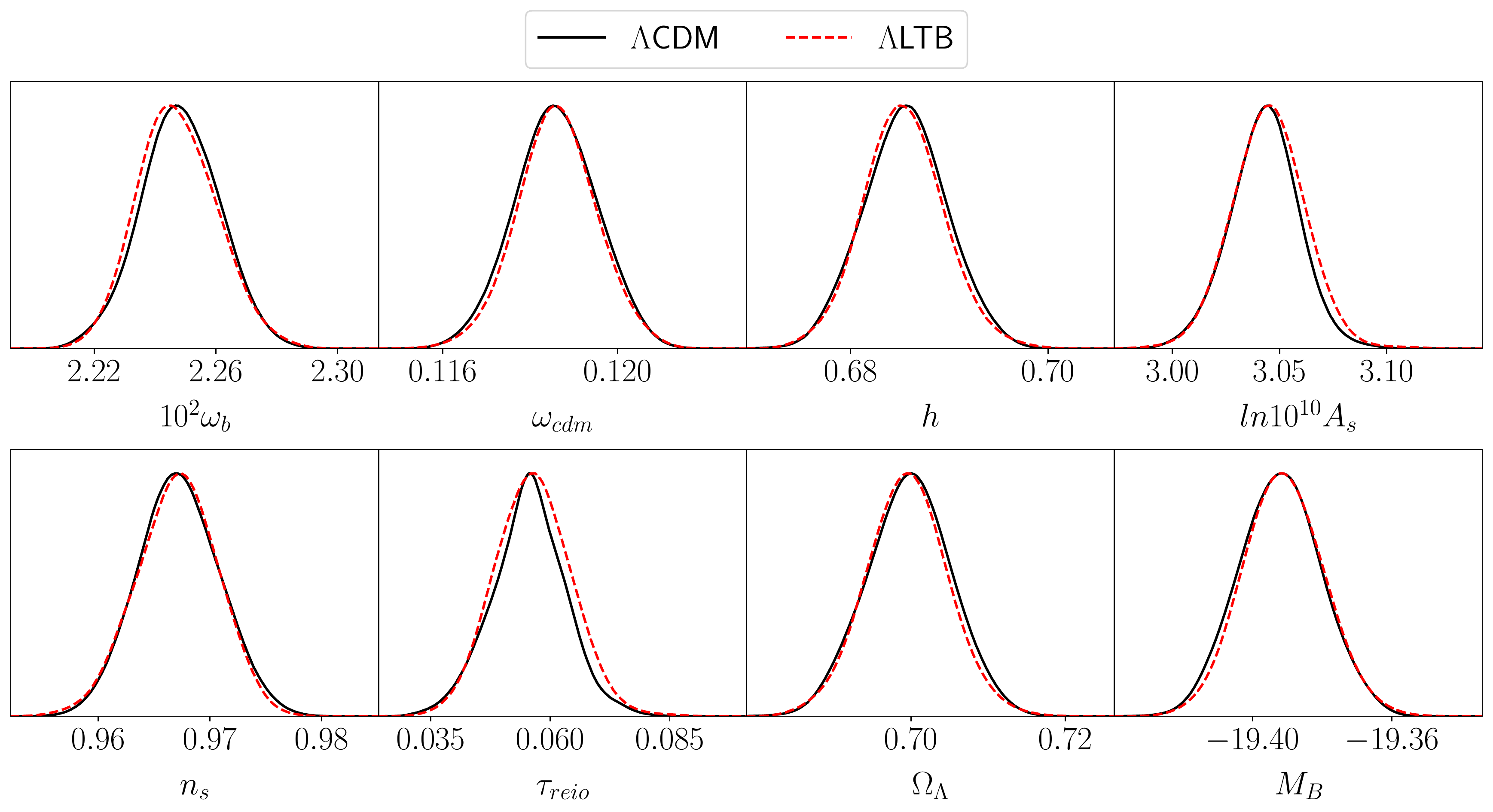}	
\caption{Marginalized constraints on the six $\Lambda$CDM parameters and also the derived parameter $\Omega_\Lambda$ and the supernova absolute magnitude $M_B$. We compare the constraints marginalized over the effect of inhomogeneities around us with the ones relative to the standard $\Lambda$CDM model that assumes the Copernican principle.
This plot shows that the standard $\Lambda$CDM results are robust against the effect of inhomogeneities, whose effect is basically negligible, see Table~\ref{tab:LLTB-tabs}.
In other words, cosmological inference without the Copernican principle not only is possible, but is affected to a very minor extent.}
\label{fig:LCDM-constraints}
\end{figure*}

Figure~\ref{fig:LCDM-constraints} and Table~\ref{tab:LLTB-tabs} show the constraints on the six $\Lambda$CDM parameters, marginalized over the effect of inhomogeneities around us.  For comparison sake, we also show the constraints relative to the standard $\Lambda$CDM model that assumes the Copernican principle.
Our results show that dropping the Copernican principle has a minor effect on the $\Lambda$CDM parameters, slightly increasing the  allowed parameter region because of the small correlations with the $\Lambda$LTB parameters. We show in Appendix~\ref{ap:triplot} the triangular plot with all the correlations.

\begin{table}
\begin{center}
\setlength{\tabcolsep}{15pt}
\renewcommand{\arraystretch}{1.35}
\begin{tabular}{lcc}
\hline
\hline
Parameter           & $\Lambda$CDM & $\Lambda$LTB \\ \hline
$10^{2}\omega_{b }$ & $2.25^{+0.026}_{-0.027}$ & $2.25^{+0.027}_{-0.025}$ \\ \hline
$\omega_{cdm }$ & $0.119^{+0.002}_{-0.002}$ & $0.119^{+0.002}_{-0.002}$ \\ \hline
$H_0$ & $68.56^{+0.84}_{-0.82}$ & $68.53^{+0.82}_{-0.81}$ \\ \hline
$ln10^{10}A_{s }$ & $3.04^{+0.03}_{-0.03}$ & $3.04^{+0.03}_{-0.03}$ \\ \hline
$n_{s }$ & $0.967^{+0.007}_{-0.007}$ & $0.967^{+0.007}_{-0.007}$ \\ \hline
$\tau_{reio }$ & $0.056^{+0.016}_{-0.016}$ & $0.056^{+0.016}_{-0.015}$ \\ \hline
$\Omega_{\Lambda}$ & $0.70^{+0.011}_{-0.011}$ & $0.70^{+0.011}_{-0.011}$ \\ \hline
\hline
\end{tabular}
\caption{68\% confidence level intervals for the six $\Lambda$CDM parameters and also the derived parameter $\Omega_\Lambda$, marginalized over the effect of inhomogeneities around us ($\Lambda$LTB) and for the standard $\Lambda$CDM model that assumes the Copernican principle.}
\label{tab:LLTB-tabs}
\end{center}
\end{table}

\section{Discussion}\label{sec:discussion}

\subsection{The local structure and the $H_0$ tension}\label{sec:void}

If only the CMB, the prior on the supernova absolute magnitude $M_B$ and the low-redshift supernovae are included in the analysis, then one sees from Figure~\ref{fig:LLTB-constraints} that a local underdensity of effective size 600--900 Mpc and effective depth of -(0.2--0.1) is favored by the data, strongly at odds with the Copernican prior. This shows how a void can solve the $H_0$ tension by boosting the local Hubble rate~\citep[see][]{Camarena:2021-LLTB-H0}.
However, it is enough to include the full supernova dataset to see that the void scenario is excluded.
In other words, although the local structure may cause environmental effects such as a possible bias on the local value of $H_0$~\citep[see][and references therein]{Camarena:2018nbr}, we find that a local inhomogeneity cannot solve the $H_0$ crisis.
We discuss this thoroughly in \citet{Camarena:2021-LLTB-H0} \citep[see also][and references therein]{Cai:2020tpy}.

From Figures~\ref{fig:LLTB-constraints} and~\ref{fig:excess} it is also clear that available observations started to probe the local structure, going beyond cosmic variance expectations. 
There have been claims that we live inside a local void, see, for instance, \citet{Keenan:2013mfa,Whitbourn:2013mwa,Boehringer:2019xmx,Colgain:2019pck}. These  claims were challenged by the analyses of, e.g., \citet{Kenworthy:2019qwq,Lukovic:2019ryg}. According to our results, deep structures are allowed only on very local scales, $\lesssim 100$ Mpc, possibly contradicting the claims by \citet{Keenan:2013mfa,Whitbourn:2013mwa,Boehringer:2019xmx}, which suggest  larger voids. In particular, by comparing the observational constraints with the ones from the Copernican prior, we do not find a marked preference for underdensities with respect to overdensities.%
\footnote{Note that, as said after eq.~\eqref{deltatilde0}, we are showing results for $\tilde \delta$.}
However, one must note that, on such small scales, anisotropies play an important role, which is not captured by our modeling.
Therefore, it is not straightforward to compare our results with analyses that model local anisotropies.

\subsection{Towards inhomogeneous cosmology}\label{sec:therole}

Figure~\ref{fig:LLTB-constraints} shows how the region of the $\delta_L$-$r_L^{\rm out}$ parameter space that is allowed by data is progressively constrained to closely follow the one allowed by the Copernican prior.
It is interesting to note that while the case CMB+SNe+$M_B$ already tightly constrains the $\Lambda$LTB model, large-scale inhomogeneities at $r_L^{\rm out} \gtrsim 500$ Mpc are efficiently constrained only by the combinations of all probes, showing their synergies in constraining deviations from FLRW.
These results represent a substantial improvement as compared to the previous analysis of \citet{Valkenburg:2012td}.

Globally, one can quantify how much non-Copernican structure is allowed by comparing, in Figure~\ref{fig:LLTB-constraints}, the CP area with the one allowed by data, as proposed by \citet{Valkenburg:2012td}.
Table~\ref{tab:area-tabs} shows the ratios of the areas of the $2\sigma$ contours for the different cases here analyzed. One can note that, when the whole parameter space  is considered, the ratio is close to $1$. However, as remarked earlier, very-large-scale inhomogeneities are more difficult to constrain and so we also compute the ratios considering only scales $r^{\textrm{out}}_L \ge 190$ Mpc. These results show that the ratio is $\sim 3$ when CMB+SNe+$M_B$ are considered and decreases to $\sim 2$ when all data are included.

Finally, we also considered the case of nonzero background curvature and found that our results remained basically unaltered. The reason is that CMB strongly constrains the background value of $\Omega_k$, and this is not affected by the compensated LTB inhomogeneity, which is constrained to small contrasts by the other observables.

All these results imply that, within the present modeling, we are close to establishing the Copernican principle and, even more important, that dropping the Copernican principle assumption does not imply  worse constraints on the cosmological parameters.

\begin{table}
\begin{center}
\setlength{\tabcolsep}{3pt}
\renewcommand{\arraystretch}{1.35}
\begin{tabular}{lcc}
\hline
\hline
\multirow{2}{*}{Case} & $A_{\rm obs}$/$A_{\rm CP}$  & $A_{\rm obs}$/$A_{\rm CP}$ \\ 
    & $0 \!\leq\! r^{\textrm{out}}_L$  & $190 \text{Mpc} \!\leq\! r^{\textrm{out}}_L $ \\ \hline
CMB + SNe + $M_B$ & $1.16$ & $2.85$ \\ \hline
CMB + SNe + $M_B$ + BAO + HZ & $1.11$ & $2.88$ \\ \hline
CMB + SNe + $M_B$ + $y$-dist. & $1.12$ & $2.83$ \\ \hline
CMB + SNe + $M_B$ + kSZ & $1.07$ & $2.35$ \\ \hline
CMB + SNe + $M_B$ + All & $1.02$ & $2.15$ \\ \hline
\hline
\end{tabular}
\caption{Ratios of the areas of the $2\sigma$ constraints from observations and the Copernican principle, see Figure~\ref{fig:LLTB-constraints}.}
\label{tab:area-tabs}
\end{center}
\end{table}

\subsection{The LTB parametrization}\label{sec:par}

It is important to mention that our results depend, to some extent, to the chosen parametrization for the curvature function given in eq.~\eqref{eq:kr}.
While it is clear that the two main physical parameters describing a spherical inhomogeneity are its size $r_b$ and contrast $\delta_0$, it is nevertheless true that, with ever tighter constraints, details of the curvature profile such as its smoothness could start having an impact.
One could overcome this limitation by considering a more flexible parametrization for the curvature or density profile as proposed in \citet{Redlich:2014gga}, where a an $n$-node spline is considered. This approach is clearly recommended if one wants to find the best-fit inhomogeneous model to observations and will be pursued in \citet{Camarena:2021-LLTB-H0}, but this is not our scope here.

Here, we wish to test the Copernican principle, that is, test all the parameter space of LTB models and the adoption of a more general profile may lead to problems. Indeed, for the analysis of the present work to be meaningful, we wish to explore the overdensities and underdensities in a similar fashion and this is not a trivial task within the LTB model of eq.~\eqref{eq:metric}. The reason is that underdensities may experience shell-crossing singularities which, although unphysical, prevent the analysis and create an artificial asymmetry in the parameter space.
Shell crossing occurs when $R'=0$ and this happens when the inner faster-expanding underdensity pushes against the compensating shell.
In other words, when exploring the parameter space of a more flexible profile, shell crossing  could lead to volume effects which would bias the results.

By inspecting the distribution of models for which the computation of the LTB dynamics failed, we checked that the parametrization of eq.~\eqref{eq:kr} does not penalize underdensities or overdensities.
However, one has to keep in mind that the results shown in Figures~\ref{fig:LLTB-constraints} and~\ref{fig:excess} are conditional to the assumed parametrization of the LTB model.

\section{Conclusions}\label{sec:conclusions}

The analysis carried out in this work is but a first step in the direction of analyzing and interpreting cosmological and astrophysical data within the framework of inhomogeneous cosmologies. Inhomogeneous cosmology is loosely defined as cosmology without the assumption of large-scale isotropy and homogeneity, that is, it is not based on an {\it a prior} assumed FLRW metric.
As discussed in the Introduction, data themselves may suggest  that the universe could  feature large-scale inhomogeneities and isotropies beyond the standard model of cosmology. Consequently, it is important to pursue a program that confront observations with arbitrarily inhomogeneous cosmologies.

Here, we adopted the simple approach of endowing the $\Lambda$CDM model with a spherical inhomogeneity. We found that, within our LTB parametrization, data can tightly constrain this extra inhomogeneity.
Also, our results show that the constraints on the standard $\Lambda$CDM parameters are not weakened after marginalizing over the local structure. In other words,  dropping the Copernican principle assumption does not necessarily imply significantly worse constraints on, e.g., the dark energy density. This positive result confirms that the present and future data can be meaningfully analyzed within the framework of inhomogeneous cosmology.

A possible development of the present analysis is to consider inhomogeneities in the radiation field, as proposed by \citet[]{Regis:2012iq}. Indeed, if the universe features large-scale inhomogeneities in the matter, one may expect a similar behavior in the other fields such as the baryon fraction or baryon-to-photon ratio which can significantly alter some of the analysis and constraints. We envision that  present and future cosmological data will nevertheless be able to constrain the free functions of these models.

Finally, mapping the local structure may have important implications; a notable one is its effect on the $H_0$ crisis, which we discuss in a separate paper where we derive a robust constraint on the local value of $H_0$ from the $\Lambda$LTB model \citep{Camarena:2021-LLTB-H0}.

\section*{Acknowledgements}

It is a pleasure to thank Wessel Valkenburg for sharing \texttt{VoidDistances2020}.
DC thanks CAPES for financial support.
VM thanks CNPq and FAPES for partial financial support. CC is supported by the UK Science \& Technology Facilities Council Consolidated Grant ST/P000592/1.
This project has received funding from the European Union’s Horizon 2020 research and innovation programme under the Marie Skłodowska-Curie grant agreement No 888258. This work made use of the CHE cluster, managed and funded by COSMO/CBPF/MCTI, with   financial   support   from   FINEP   and   FAPERJ, and  operating  at  the  Javier  Magnin  Computing  Center/CBPF. 
This work also made use of the Virgo Cluster at Cosmo-ufes/UFES, which is funded by FAPES and administrated by Renan Alves de Oliveira.

\section*{Author contributions}

VM and CC conceived the research question.
All authors designed the study and analysis plan.
DC led the numerical implementation of the model and observables.
ZS contributed to the numerical implementation and MCMC exploration.
DC and VM drafted the initial version of the manuscript.
All authors critically reviewed early and final versions of the manuscript.

\section*{Data availability}

The data underlying this article will be shared on reasonable request to the corresponding author.
The \texttt{monteLLTB} code is available at \href{https://github.com/davidcato/monteLLTB}{github.com/davidcato/monteLLTB}.

\bibliographystyle{mnrasArxiv}
\bibliography{biblio}

\appendix

\section{The \lowercase{monte}LLTB code}\label{ap:code}

We embeded the \texttt{vd2020} code (available at \href{https://github.com/valkenburg/vd2020}{github.com/valkenburg/vd2020}) into \texttt{montepython} to create \texttt{monteLLTB}: a cosmological solver and sampler for the $\Lambda$LTB model. Taking advantage of the likelihood and sampler structure of \texttt{montepython} we include the $\Lambda$LTB cosmology by adapting the likelihood computation scheme. We started defining the method \verb|ini_LLTB| on \verb|sampler.py|,
which executes the solver \texttt{vd2020} considering the current sampled point. Then, a call for \verb|ini_LLTB| is included into the method \verb|compute_lkl| to pass the $\Lambda$LTB solution to the corresponding likelihood. Note that this is possible since the method of the likehood \verb|loglkl| now receives a new argument \verb|LLTBin|, which contains the $\Lambda$LTB solution. We also modified the likelihoods in order to compute the observables according the $\Lambda$LTB predictions. Note that the output of \texttt{vd2020} is managed by the file \verb|LLTB_functions.py|, which contains definitions of distances and metric functions. 
Finally, it is important to mention that we modified \texttt{vd2020} in order to customize the management of error, output precision and outputted functions.
However, the core of the $\Lambda$LTB solver, the implementation to compute $R(t,r)$ through Carlson’s elliptic integrals \citep{Valkenburg:2011tm}, remained unchanged. The \texttt{monteLLTB} code is available at \href{https://github.com/davidcato/monteLLTB}{github.com/davidcato/monteLLTB}.

\section{Impact of large scales}\label{ap:lowl}

\begin{figure}
\centering 
\includegraphics[width=1.05 \columnwidth]{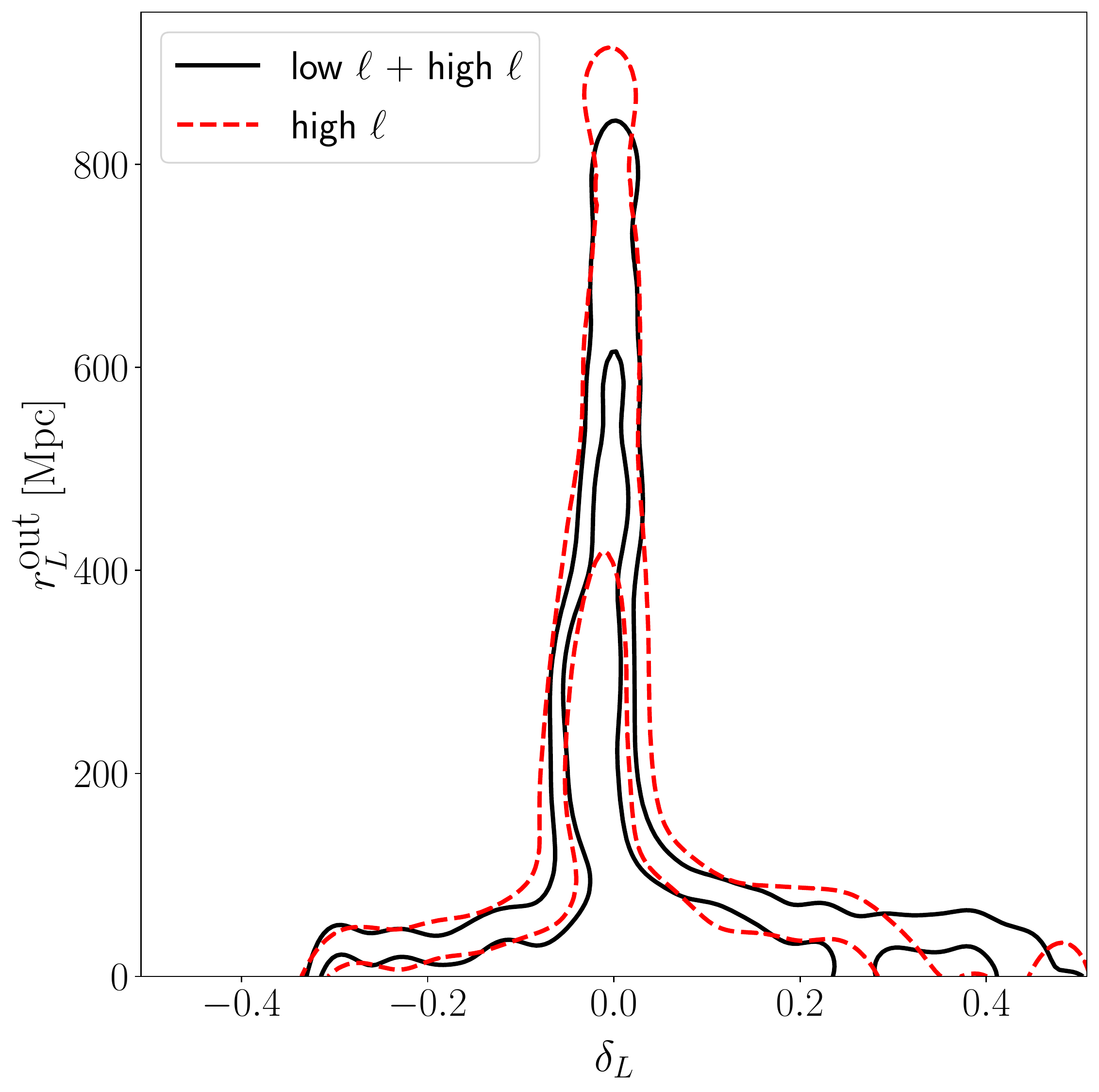}
\includegraphics[width=1.05\columnwidth]{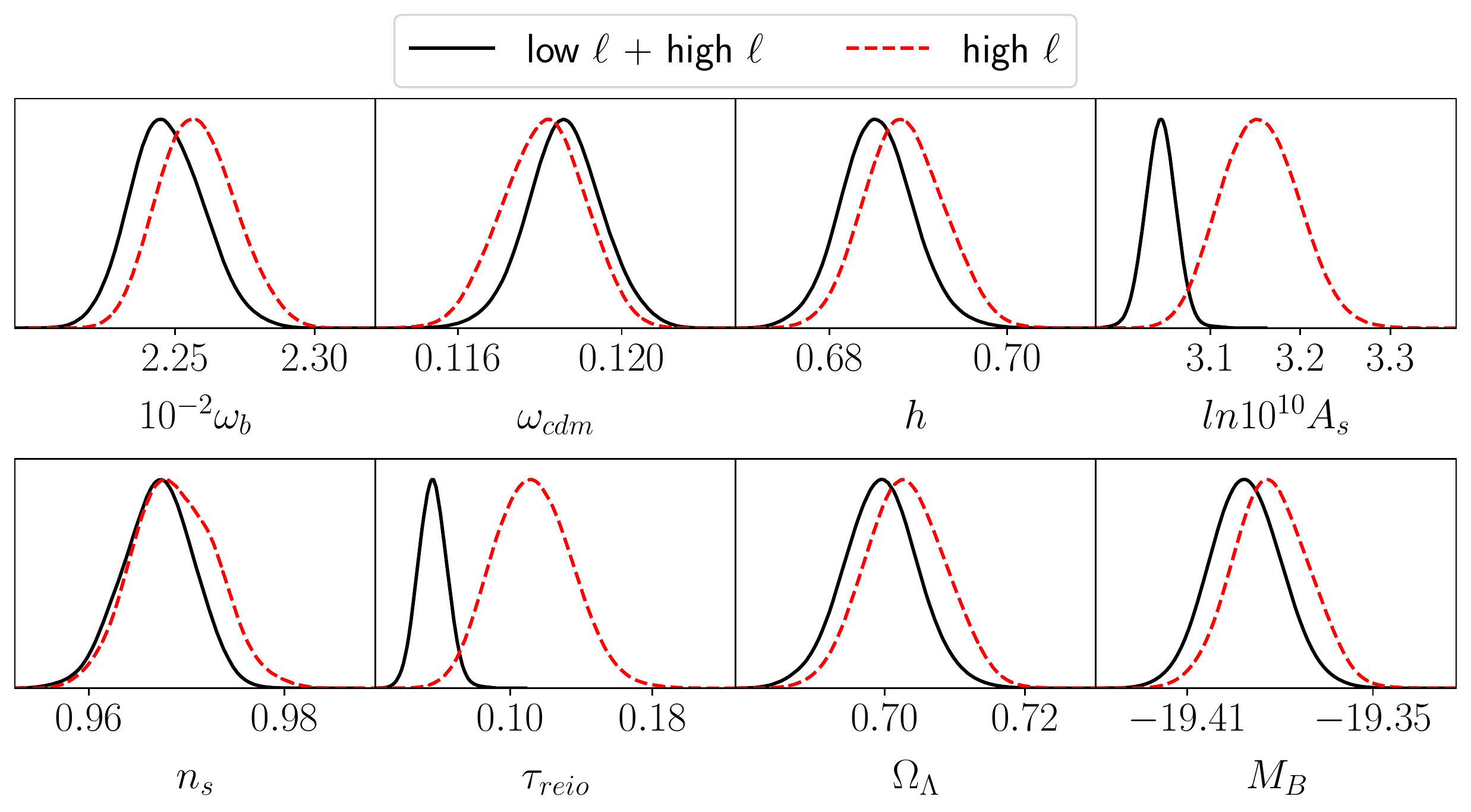}
\caption{Marginalized constraints with/out low-$\ell$ Planck data.}
\label{fig:LLTB-lowl}
\end{figure}

Here, as discussed in Section~\ref{sub:CMB}, we assess the impact of not using low-$\ell$ Planck data.
Figure~\ref{fig:LLTB-lowl}
compares the constraints when using both high- and low-$\ell$ Planck data with the more conservative case of only including high-$\ell$ Planck data. We see that the impact on the parameters of the inhomogeneity is minor, while the impact on the $\Lambda$CDM parameters is, as expected, strong. In other words, in the present analysis, the low-$\ell$ Planck data are effective only for the $\Lambda$CDM parameters.
A more complete treatment requires the challenging computation of  perturbations in an inhomogeneous background.

\section{Triplot}\label{ap:triplot}

\begin{figure*}
\centering 
\includegraphics[width=\textwidth]{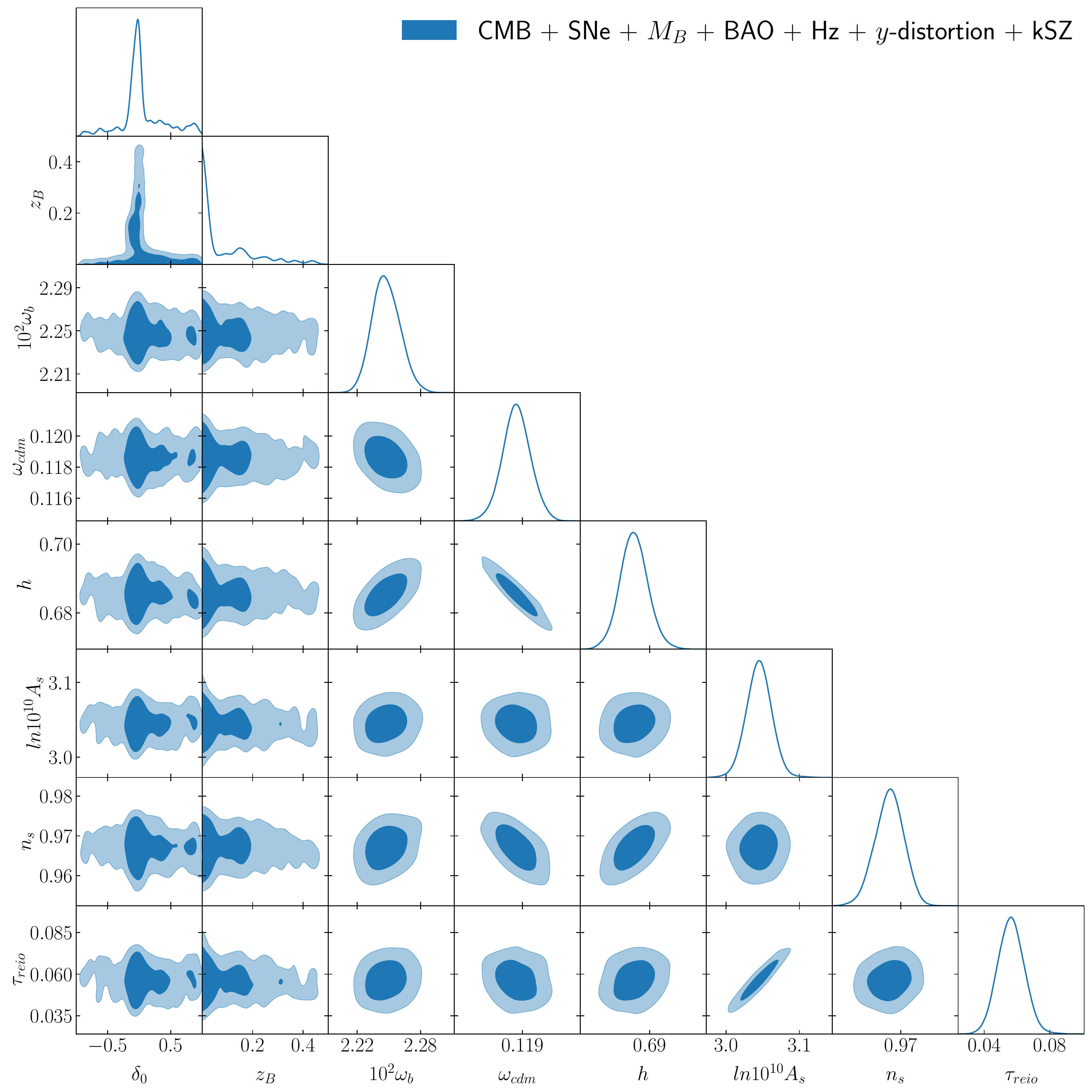}
\caption{68\% and 95\% marginalized constraints on the eight independent parameters of the $\Lambda$LTB model.}
\label{fig:triplot}
\end{figure*}

For completeness we show in Figure~\ref{fig:triplot} the marginalized constraints and correlations of the eight independent parameters of the $\Lambda$LTB model.

\bsp	
\label{lastpage}
\end{document}